\documentclass[preprint,12pt]{elsarticle}

\usepackage{graphicx}
\graphicspath{ {pictures/} }
\usepackage{amssymb}
\usepackage{amsmath}

 \usepackage{float}

\journal{Intel STS 2015}

\begin{document}

\begin{frontmatter}


\title{Development of a Computationally Optimized
Model of Cancer-induced Angiogenesis through
Specialized Cellular Mechanics}



\author{Dibya Jyoti Ghosh}

\address{California, United States}

\begin{abstract}
Angiogenesis, the development of new vasculature, is a critical process in the growth of new tumors. Driven by a goal to understand this aspect of cancer proliferation, we develop a discrete computationally optimized mathematical model of angiogenesis that specializes in intercellular interactions. We model vascular endothelial growth factor spread and dynamics of endothelial cell movement in a competitive environment, with parameters specific to our model calculated through Dependent Variable Sensitivity Analysis (DVSA) and experimentally observed data. Through simulation testing, we find the critical limits of angiogenesis to be 102 µm and 153 µm respectively, beyond which angiogenesis will not successfully occur. Cell density in the surrounding region and the concentration of extracellular matrix fibers are also found to directly inhibit angiogenesis. Through these three factors, we postulate a method for establishing criticality of a tumor based upon the likelihood of angiogenesis completing. This research expands on other work by choosing factors that are patient-dependent through our specialized Cellular Potts mode, which serves to optimize and increase accuracy of the model. By doing such, we establish a theoretical framework for analyzing lesions using angiogenetic properties, with the ability to potentially compute the criticality of tumors with the aid of medical imaging technology.
\end{abstract}


\end{frontmatter}


\section{Introduction}

Over 7 million people die every year of cancer, killed by false diagnoses, improper treatments, and excessive medications. Although many aspects of cancer proliferation and detection have been documented over the years, one newly emerging category is the field of angiogenesis. Angiogenesis is the development process of new blood vessels in the cardiovascular system to get oxygen and nutrients to all systems in the body, a process that drives cancer proliferation. Research on cancer-induced angiogenesis focuses on the relationship between blood vessels and cancerous lesions, detailing the mutual growth in the interdependent processes. Mathematical models of angiogenesis can simulate the system to analyze stresses on the tumor, but current models are very theoretical with minimal practical applications. This paper focuses on patient-based factors, borrowing from previous models, to create a fully integrated system. In the process, I theorize a high level specialized mechanism for cellular mechanics that optimizes computation and accuracy in the system.

\section{Biological Background}
\label{S:1}
\subsection{Angiogenesis}
Angiogenesis, the development of new blood vessels from existing vasculature, is an integral part of many physiological processes in the human body, including wound healing, tumor growth and diabetes. Growth factors trigger a biochemical pathway in the blood vessels, causing hemangioblasts in the system to specialize into vascular endothelial cells.(Karamysheva, 2008). These new endothelial cells form a new blood vessel with branches off from the existing capillary and migrates through chemotaxis toward the source of these growth factors.\cite{Smith:2012qr} \\
\subsection{Cancer Induced Angiogenesis}
~\\
Cancerous cells, like all other cells, require basic nutrients and oxygen to grow and proliferate. Yet due to the high oxygen consumption of cancerous cells, the tumor is often hypoxic (lack of oxygen) and its growth stagnates. While in this hypoxic state, tumors begin secreting growth factors that are triggers for angiogenesis (Nishida Yano, 2006). While in hypoxia, the tumor is at its critical limit, where the vascular response to the secreted chemicals directly affects the lifespan of the tumor. If angiogenesis occurs, then a blood vessel is attached very close to the tumor, allowing it to expand and perhaps metastasize. If angiogenesis fails to occur, then the tumor will continue subsisting and not growing until the immune system triggers forced apoptosis on the cell. Angiogenetic growth factors will continue flowing, until the tumor receives negative feedback through oxygen. \\
\subsection{Vascular Endothelial Growth Factor} ~\\
VEGF ( Vascular Endothelial Growth Factor) is a family of homodimeric glycoproteins known to trigger angiogenesis and induce vascular permeability in endothelial cells. Previous studies have shown that VEGFA is the primary growth factor secreted by tumors when deprived of oxygen (Gabhann Popel, 2003). Endothelial cells have multiple receptors for the VEGF family, VEGFR1, VEGFR2, and VEGFR3. Of theses, only VEGFR2 initializes the biochemical process for angiogenesis, but the other members of the VEGFR family compete for VEGFA, resulting in a complex chemical system. Once the new vasculature is formed, the tip cell of the blood vessel moves up the concentration gradient of VEGF towards the tumor. \textit{(Unless specifically mentioned, VEGFR will refer to VEGFR2, and VEGF and Φ will refer to VEGFA)}\

\section{Tumor Angiogenesis Modelling}
\par
A sudden surge in biological research in angiogenesis in the late 20th century has triggered many mathematical efforts to model the process. Initial simulations were continuous models focused either on tip cell dynamics for endothelial cells (Szabó) or biochemical interactions with growth factors (Gabhann Popel, 2003) . Following initial combinatory models which attempted to link the dynamical and chemical aspects, the first complete discrete model of angiogenesis came out in 2006 (Norrby, 2006). Later on, further research expanded the two aspects, as well as using a 3D environment, taking the first steps to realistic simulations of cancer-induced angiogenesis.
\par
Our model builds upon the model of Zheng, with an additional emphasis on cellular mechanics. In line with the goal for integration of this model with medical technology, we develop a discrete dynamic system to analyze all aspects of the environment. The two primary segments of this model are the cellular environment and the chemical system that inhabits it. For the chemical aspect, we focus on biochemical interactions like VEGF uptake by endothelial cells and intra-chemical interactions like diffusion, dissolution, and decay. When examining forces that influence the cellular environment, we theorize that the normal noncancerous cells surrounding the tumor play a distinctly important role in the probability of angiogenesis occurring. Thus, our model develops an adaptation of the Cellular Potts cellular automata model that places special emphasis on the normal cell surrounding the tumor.
\par
We integrate the biochemical model and the cellular mechanics model together on a multi-scale level. Although these models disagree on temporal and spatial scales, we manipulate the models to sacrifice accuracy in exchange for a well-integrated and more computationally efficient model. The environment we design is a 3d collection of points, with the tumor located upon the upper right corner of the environment, and blood vessels orthogonal to the X and Y axes through the origin (Figure 1). Since cellular growth mechanics are anomalous in cancerous cells, we place the tumor outside the system to prevent interference in the model. The blood vessel in question is also designed to be outside the environment, and intersecting the environment at the origin, where angiogenesis will take place. By manipulating the environment, through placement of the blood vessel and tumor, we prevent aberrant results and increase future reproducibility.
\begin{figure}[h]
\centering
\includegraphics[width=0.9\textwidth]{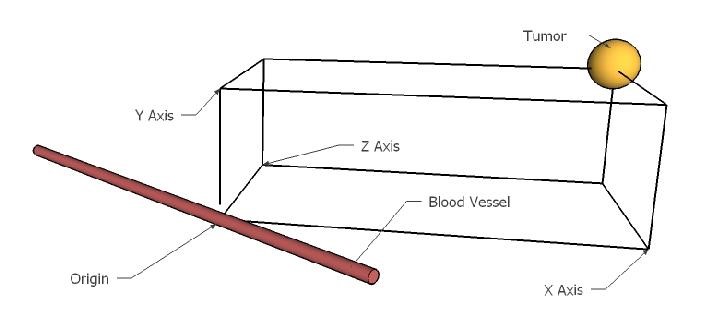}
\caption{Location of critical components in the cellular environment for the model}
\end{figure}
~\\~\\

\section{Biochemical Model of VEGF Interactions}\
\subsection{Secretion of VEGF}
We describe the level of VEGF being secreted from the tumor in a hypoxic environment through its distance from the blood vessel. Having established that all the body cells in the environment are not hypoxic, we reduce complexity of the system by substituting the concentration of the oxygen with r, the distance of the tumor from the blood vessel. A rudimentary diffusion substitution between oxygen and radius yields Equation 1. The secretion of VEGF occurs only from the center of the tumor, a point source, with Thus, the VEGF secretion will decrease to zero as angiogenesis proceeds.
\begin{align}
\frac{d\phi}{dt} &= \frac{k_s}{r^3}
\end{align}
\subsection{Interchemical Interactions}
We model typical chemical interactions outside cells such as diffusion and decay, which affect spread of the growth factor. Borrowing from previous research (Gabhann Popel, 2003), this model simulates a linear decay of VEGF. Equations 2 and 3 represent the continuous differential equation and partially-solved PDE versions of the decay equation respectively. We simplify the PDE by connecting it to the previous time step, thus discretizing the variable and increasing computational efficiency. (Equation 4).
\begin{align}
\frac{d\phi}{dt} &= -k_{dec}\phi \\
\phi(t) &= \phi_ie^{-\phi t} \\
\phi(t) &= \phi(t-1)e^{-\phi}
\end{align}
Previous research in VEGF modelling (Jain, 2008) described VEGF diffusion in only 1 dimension. To increase accuracy and link the simulation to a 3D environment, we implement a 3D model of diffusion, radiating from the point source of VEGF. We derive Fick’s law to develop a mechanism to model concentration changes over time. Our model of VEGF diffusion is based upon Fick’s first (Equation 5) and second laws (Equation 6), where J represents the flux of chemical movement, and D is the diffusion constant. We deviate from Fick’s continuous law by establishing a variable concentration gradient, and so we must discretize diffusion using the previous time step. In this new equation (Equation 7), the specified location is categorized as a receiver of diffused chemicals, with the 6 neighboring cells as the sources of chemicals.
\begin{align}
J &= -D\nabla\phi \\
\frac{\partial \phi}{\partial t} &= D\frac{\partial^2 \phi}{\partial t^2}\\
\phi(t) &= \phi(t-1) - \frac{1}{6}D\sum_{n=1}^{6} \nabla\phi_{n}
\end{align}
\subsection{VEGF Pathway for Angiogenesis}
VEGF and its complement receptor family VEGFR are the gateway to hemangioblast specialization into endothelial tissue. Unlike previous research, which model just the VEGF-A proteins and VEGFR2 receptors, this model extends to VEGF-B and VEGFR1. These chemicals do not trigger angiogenesis but they indirectly inhibit the target reaction.
\begin{align*}
VEGFA + VEGFR1 & \overset{k_1}{\rightarrow} Complex_{VEGFR1} \\
VEGFB + VEGFR1 &\overset{k_2}{\rightarrow} Complex_{VEGFR1}\\
VEGFA + VEGFR2 &\overset{k_3}{\rightarrow} Complex_{VEGFR2}\\
VEGFB + VEGFR2 &\overset{k_4}{\rightarrow} Complex_{VEGFR2}\\
Complex_{VEGFR1} &\overset{k_5}{\rightarrow} VEGFR1 \\
Complex_{VEGFR2} &\overset{k_6}{\rightarrow} VEGFR2 \\
\end{align*} 

Applying the law of mass action to this set of chemical reactions gives us the rate of cellular use of VEGF-A. Using the steady-state solution obtained from Michaelis-Mentin kinetics from Figure 2, we can find the exact timing of successful angiogenesis.
\begin{align}
\frac{\mathrm{d}[VEGFA]}{\mathrm{d} t} &= -k1[VEGFA][VEGFR1]\\
\frac{\mathrm{d}[VEGFR1]}{\mathrm{d} t} =& -k1[VEGFA][VEGFR1] \\ 
&-k2[VEGFB][VEGFR1]+k5[{Complex}_{VEGFR1}] \notag
\end{align}
\subsection{State Function and Integration}
Integration of the biochemical and the mechanical models occurs by assuming the same physical and temporal scale. Resolution is unimportant, since concentration level will not greatly differ, so we lose minimal accuracy, yet by linking the two models, and speed it up by at least 3 times. This thus gives way to an overall state model which allows us to find the new VEGF concentration after a single time step in 1 calculation.

\begin{align}
\frac{\partial \phi}{\partial x} &= \frac{k_s}{r^3} - k_{dec}\phi- D\nabla\phi- k_1[VEGFA][VEGFR1]
\end{align}

\section{Dynamic Mechanical Model of Intercellular Interactions}
\par
We utilize the Cellular Potts model to describe the cellular environment, a collection of cells floating in the interstitial fluid alongside the extra matrix fibers. Unlike previous cellular automata models (Bauer, May 2007) , this lattice model has only been established in the literature, and never utilized. The techniques we use to implement a Cellular Potts model are derived from similar methods in other cellular automata and implementations of other processes using this algorithm.
\par
The environment is defined as a 3 dimensional set of lattice points, all of which are assigned a non-unique Cell ID, defined as µ. This cell ID is used to differentiate between the extracellular matrix, the interstitial fluid, and the individual cells. Temporal updates to the lattice site are made upon the tendencies of the system are to shift toward minimal potential energy and maximum entropy. Every time step, though Monte-Carlo simulation, the algorithm loops through pairs of neighboring lattice points, and virtually switches all the cell IDs to update. In accordance to the Boltzmann probability distribution, we determine whether the transaction is accepted or rejected (Equation 11). Should the energy increase in the system, the change is sometimes accepted to ensure that a false equilibrium doesn’t prevent the simulation from completing angiogenesis.
\begin{align}
P_{acceptance}(\Delta E) = e^{\frac{-\Delta E}{t}}
\end{align}
\par 
In this Cellular-Potts model, we analyze four key aspects of cellular interactions, the elasticity of the cell, and cohesion between cells, cellular senescence, and chemotaxis (for endothelial cells)
\subsection{Cellular Elasticity}
Cellular elasticity refers to the potential energy driving expansion and contraction of the cell. We look at the three most important aspects of cellular elasticity in this model, volume elasticity, surface elasticity and the diffusion based elasticity. A cell that is growing will expand to the optimal size, and the physical definition of elasticity due to volume expansion says potential energy is proportional to the cube of the radius of the cell (Equation 12). Surface area, which dictates diffusion rate and thus metabolism, is also taken into consideration. Potential energy from surface area tension (Equation 13) is proportional to the square of difference in surface areas. One novel aspect that we model is the surface area to volume ratio, which dictates relative diffusion, and plays a role in correcting cell shape. Should the diffusion rate be inhibited by cell shape, then the cell tends toward its optimal configuration, a sphere. We model this as a piecewise function (Equation 14) which acts only when the surface area to volume ratio reaches a threshold.
\begin{align}
E_{Volume} &= k_{v} (v_{current}-v_{target})^3 \\
E_{SA} &= k_{sa} ({SA}_{current}-{SA}_{target})^2
\end{align}
\begin{align}
   E_{tension} = \left\{
     \begin{array}{lr}
       0 & :  \frac{1}{2}< \frac{Volume}{SA} < 2 \\
       k_{t}*{(\frac{Volume}{SA})}^4 & :  \frac{Volume}{SA} < \frac{1}{2} or \frac{Volume}{SA} > 2  \\
     \end{array}
   \right.
\end{align}
\subsection{Chemotaxis with VEGF and CXCLS}
Studies show that VEGF triggers endothelial cell migration up a concentration gradient. Empirical evidence suggests that the speed at which the cell migrates is proportional to the chemical gradient (Equation 15), with the correlation coefficient being the sensitivity of the cell. Since intercellular forces among the inner endothelial cells negate the effects of chemotaxis, this chemotactic force acts most upon the tip cell which interacts with the extracellular matrix and provides much of the direction guiding for the new vessel being formed.
\begin{align}
E_{chemotaxis} = k_{chemo}\nabla\phi 
\end{align}
\subsection{Cellular Cohesion}
We take into consideration cohesive bonds formed between the cells in the system. For the purpose of this simulation, we disregard the cohesive forces inside a cell, and also the viscosity between the interstitial fluid and the matrix fibers. Since all of the objects in the system have approximately the same density, and the lattice points are isochoric, and so we develop the mass of the system into the friction coefficients. We model the cellular cohesion is the sum of the coefficients of its neighbors which don’t belong to the same cell (Equation 16).
\begin{align}
E_{cohesion}(i) = \sum_{j=1}^{6} c_{(\mu(i),\mu(j)))}
\end{align}
\subsection{Cellular Senescence}
One major flaw that existed with previous models is the absence of cellular reproduction and cellular senescence. The surrounding environment of cells was static, and created discrepancies from physical angiogenesis experiments. The time-scale of angiogenesis is long enough for all cells to reproduce at least two times and for ~20 percent of the cells to die. We establish that a cell reproduces approximately every 20 time steps, and that cellular senescence has a 5\% chance of occurring for a cell. The probability model generated for cellular reproduction
(Equation 17) ensures that 95\% cells do reproduce within 15-25 time steps. The probability model generated for cellular senescence (Equation 18) ensures that 95\% cells will die within 1030 reproductions.

\begin{align}
P_{reproduction}(t) &= \frac{1}{2}[1+ erf(t-t_r)] \\
P_{reproduction}(t) &= \frac{1}{2}[1+ erf(\frac{t-t_r*t_s}{25\sqrt{2}})]
\end{align}

\section{Model Testing Parameters}
The mathematical model was tested in a simulation created in C++, with a parallelized structure for optimized DVSA (Dependent Variable Sensitivity Analysis). Numerical analysis is done through a structure analyzer which compares cell shape, size, and structure. Graphical analysis uses 3d space-filling models of the lattice structure, which allows for easy analysis of endothelial cell migration.
\par
The minor disconnect between the modified Cellular Potts model and the biochemical model are resolved by analytically solving the VEGF state equation through Runge-Kutta algorithms (San Joaquin Delta College), an efficient method for solving these partial differential equations.
\par
Research upon cancer angiogenesis modeling has experimentally uncovered the values for many constants used in this mathematical model. Samples include diffusion rate, decay rate of VEGF, and the relative adhesion constants between the cells. Yet, many constants such as relative tensile strength of the cells and chemotactic sensitivity to VEGF to endothelial cells, have not been developed and require a sensitivity analysis. By performing this sensitivity analysis, we find the optimal value of the constant in question and also the effect of slight change of the constant upon the system. To do such we utilize regression-based analysis, due to the high computational cost of running the operation.
\begin{table}[h]
\centering
\begin{tabular}{l l l}
\hline
\textbf{Name} & \textbf{Value} & \textbf{Equation}\\
\hline
$k_s$ & 1.531 & 1 \\
$k_dec$ & 0.011552453 & 2\\
$D$ & 1 & 5 \\
$[VEGFR]_{init}$ & 125.6 & 8\\
temperature ($t$) & 1000 & 11 \\
$k_v$ & 0.4 & 12\\
$k_{sa}$  & .4 & 13 \\ 
$k_t$ & 1 & 14 \\
$k_{chemo}$ & .15 & 15 \\
$t_r$ & 20 & 17 \\ 
$t_s$ & 20 & 17,18 \\
\hline
\end{tabular}
\caption{Constants for the simulation derived from previous research for our discretized variables }
\end{table}

\section{Computational Results and Discussion}
To test this mathematical model, we ran the model through multiple simulations. Particular factors manipulated that were of interest included cohesive strength, VEGF Sensitivity of endothelial cells and the level of migration for cells. Furthermore, to ensure that the simulation was following proper procedure, we monitor the body cell movement and VEGF concentration throughout the process. While performing these tests, we obtain the data from three major locations: a visual examination, the raw variable stream from the simulation, and the result on whether angiogenesis completed or not.
\subsection{Cohesion Strength}
In general, manipulation in the cohesive strength alters the shapes of cells and the distance between cells. Typically, body cells will have similar cohesion constants, and not much variance will occur. However as seen by tip cell research, the cohesion coefficients between the endothelial cells and interstitial fluid will vary from location to location.
\par
\begin{figure}[H]
\centering
\includegraphics[width=0.4\textwidth]{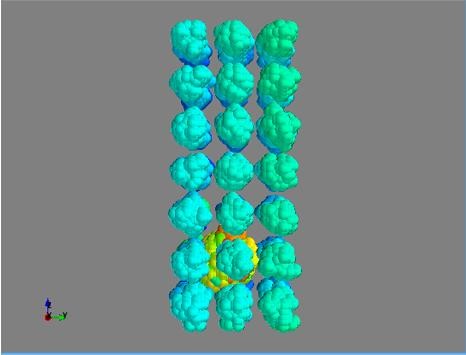}
\includegraphics[width=0.4\textwidth]{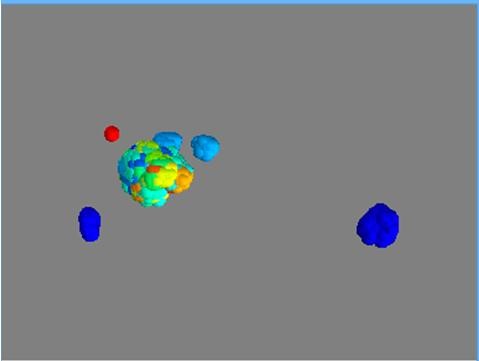}
\caption{Space filling model for simulation with high cohesion forces. The entire angiogenesis environment is displayed on the left, and just the new vasculature on the right}
\end{figure}

\par
Our testing shows that when the cohesion coefficient is extremely high (Figure 3), clumping occurs and the space repression factors overpower the force of chemotaxis. Such is remedied in two manners, either by triggering cancer angiogenesis in a neighboring vasculature or by adjusting the environment by secreting extracellular matrix fibers.

\subsection{VEGF Sensitivity}
Beyond the initial biochemical pathway that triggers angiogenesis, the main factor driving endothelial cells is VEGF chemotaxis on endothelial cells. Thus, this is not a factor that can be manipulated in the real world. However, by analyzing how VEGF causes cells to move helps us calibrate the proper sensitivity to ensure that angiogenesis actually occurs. When the sensitivity is too high, all the endothelial cells move on the concentration gradient, but cause little, if any, motion for body cells. The endothelial cell begins disregarding its elastic constraints and deforms into odd shapes. When VEGF sensitivity is low, the endothelial cell begins acting like a normal body cell. Overall migration is limited to 2-3 micrometers and the new vasculature looks like a stub. The most optimal point is where the each endothelial cell migrates 20 um, and the width of the blood vessel is 10 micrometers in diameter. At this optimal point, the body cells shake around in their place, moving randomly, creating no migration power. The cellular movement due to VEGF is optimal for endothelial cells at VEGF sensitivity in Table 1. (Figure 4)
\begin{figure}[H]
\centering
\includegraphics[width=0.8\textwidth]{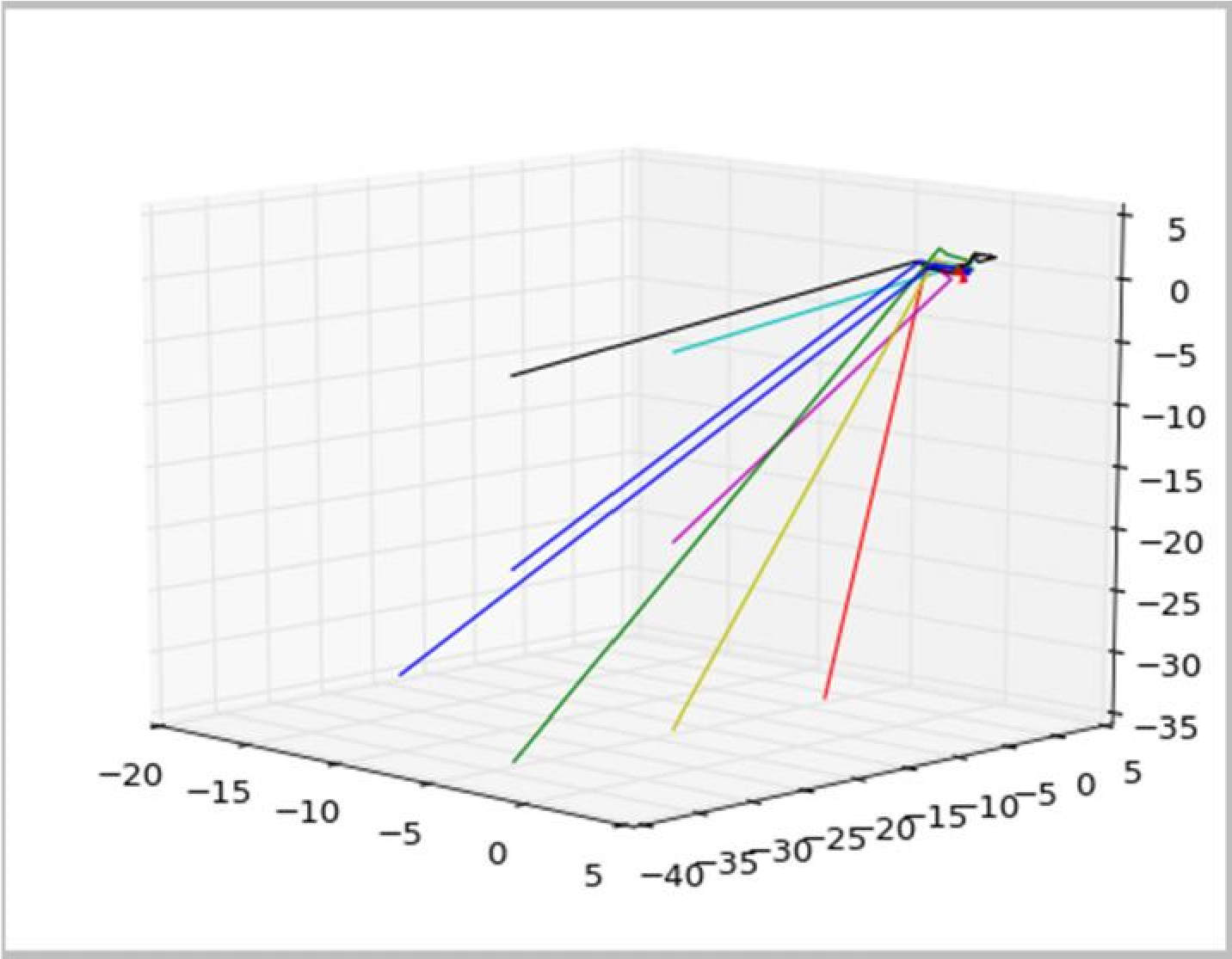}
\includegraphics[width=0.8\textwidth]{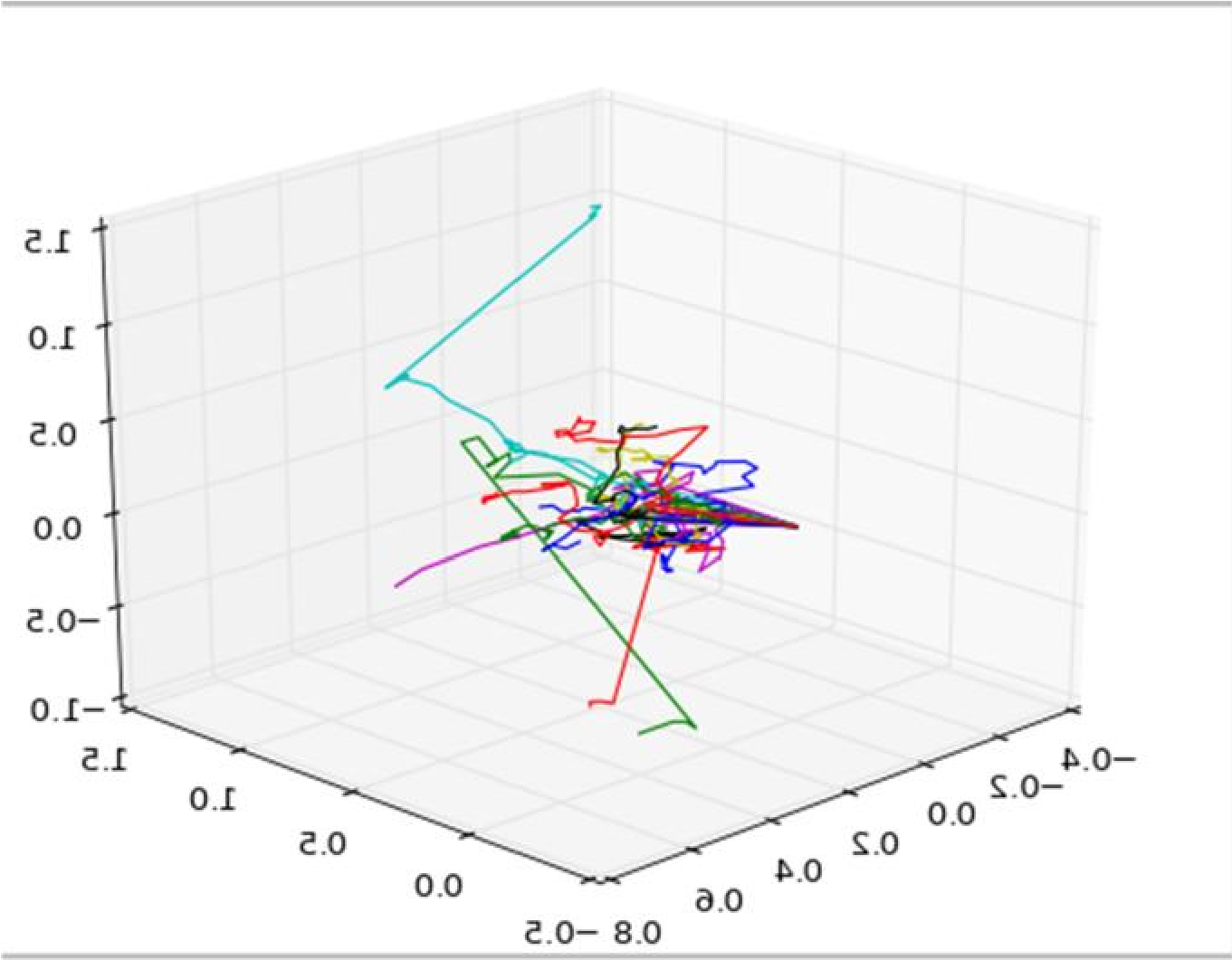}
\caption{ Cell movement for endothelial cells (left) and for body cells (right)}
\end{figure}

\subsection{Cell Density}
With our increased emphasis on the body cells, we discuss the effect of randomness of inputs on the system. Although for simulation purposes, a rectangular grid of evenly shaped and spaced out cells proves useful for reproducibility, it isn’t effective for real-life situations.
Understanding the high entropy around the body, we randomize the environment, and collect data on angiogenesis completion. The most common source of randomness is a rearrangement of the surrounding cells and matrix fibers among the interstitial fluid. We examine the effect of randomizing the environment upon the rate of angiogenesis. This change in environment does alter the rate of cancer angiogenesis by as much as 50\%, and in some cases inhibited angiogenesis. When we double the size of the initial cells and their optimal sizes, thus making the beginning cells take up significantly more room, the likelihood of angiogenesis decreases by almost 30\%. Interestingly, when the cell size is manipulated, the time taken for angiogenesis doesn’t change much, yet the probability of angiogenesis occurring in a reasonable time period decreases significantly. If cellular density is too high due to increased sizes, then initial rate of tip cell migration is affected, and probability of angiogenesis completion is reduced. However, if the tip cell does maintain a stable initial rate, then the cell density plays only a minor role in angiogenesis speed.
\subsection{Body Cell Movement}
This data shows the tracking of cell size as we progress through cancer angiogenesis. In this particular simulation, all the cells were located in lattice structure, and cell sizes were monitored for external stimuli. The increase in cell size is attributed to a force applied by the endothelial body, and the decrease in cell size, a movement away from the cell. In terms of the cellular environment, we see a pressure being put upon the side of the cube by the newly emerging endothelial cell. Although most of the cells stabilize after the first reproduction (after 20 time steps), the cells near the side of the lattice, and the size of the cell increases a result.
\begin{figure}[H]
\centering
\includegraphics[width=0.9\textwidth]{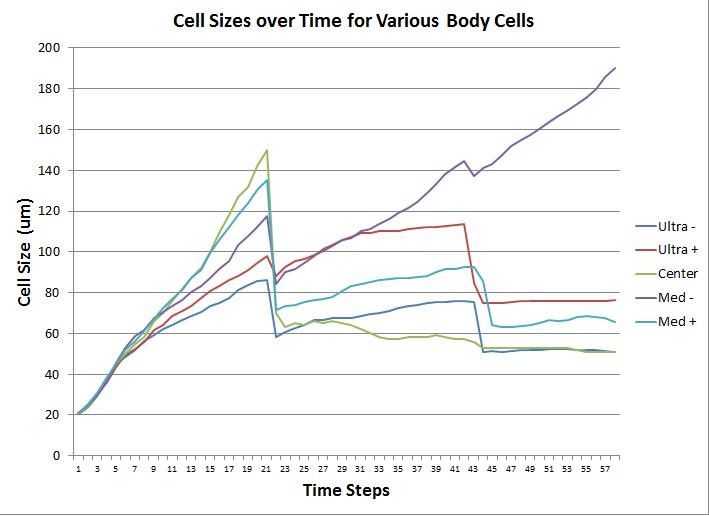}

\caption{ Results of a simulation that tracked body cell sizes to monitor the movement of endothelial cells throughout the environment}
\end{figure}

\subsection{Cell Location}
Clearly, the location of the tumor with respect to the vasculature makes a huge difference. We tested an experiment that modified the size of the environment (and thus the location of the cell), and measured the difference in time taken to complete angiogenesis. The model was calibrated with minimum distance for angiogenesis to occur at 102 um (Alarcon, 2006), and as mentioned earlier, we analytically calculated the stopping point of angiogenesis at approximately 153 um (with optimal variables and dependent upon initial cell configuration. Beyond this range, the rate of decay and diffusion reduce the supply of growth factor to the point where chemotaxis cannot drive angiogenesis.
\par
After testing for correlations between speed of angiogenesis and the distance related, we extrapolate the following relationship.

\subsection{VEGF Distribution}
VEGF is the driving factor in the entire environment, and so we observe the concentration of the chemical as angiogenesis takes place. Initially, after the tumor begins experiencing hypoxia, the concentrations increase in the area of the tumor, proving that the tumor is properly secreting VEGF. VEGF begins to spread and begins reaching appreciable quantities in the location of the blood vessel. Yet, due to the rapid rate of reaction for the angiogenesis pathway for endothelial cells, the difference cannot be seen until the angiogenesis pathway completes and the reaction stops occurring in the region. At that point, we see the VEGF concentration shoot up. This acts as a partial deterrent to the VEGF chemotaxis, but to overcome the initial spike in concentration, we notice that the tension force applied by the endothelial cells. As the angiogenesis completes, we reach equilibrium, and the overall VEGF gradient present removes. We ensure the veracity of this VEGF proliferation model by comparing these trends with previous research (Norrby, 2006).
\begin{figure}[H]
\centering
\includegraphics[width=0.9\textwidth]{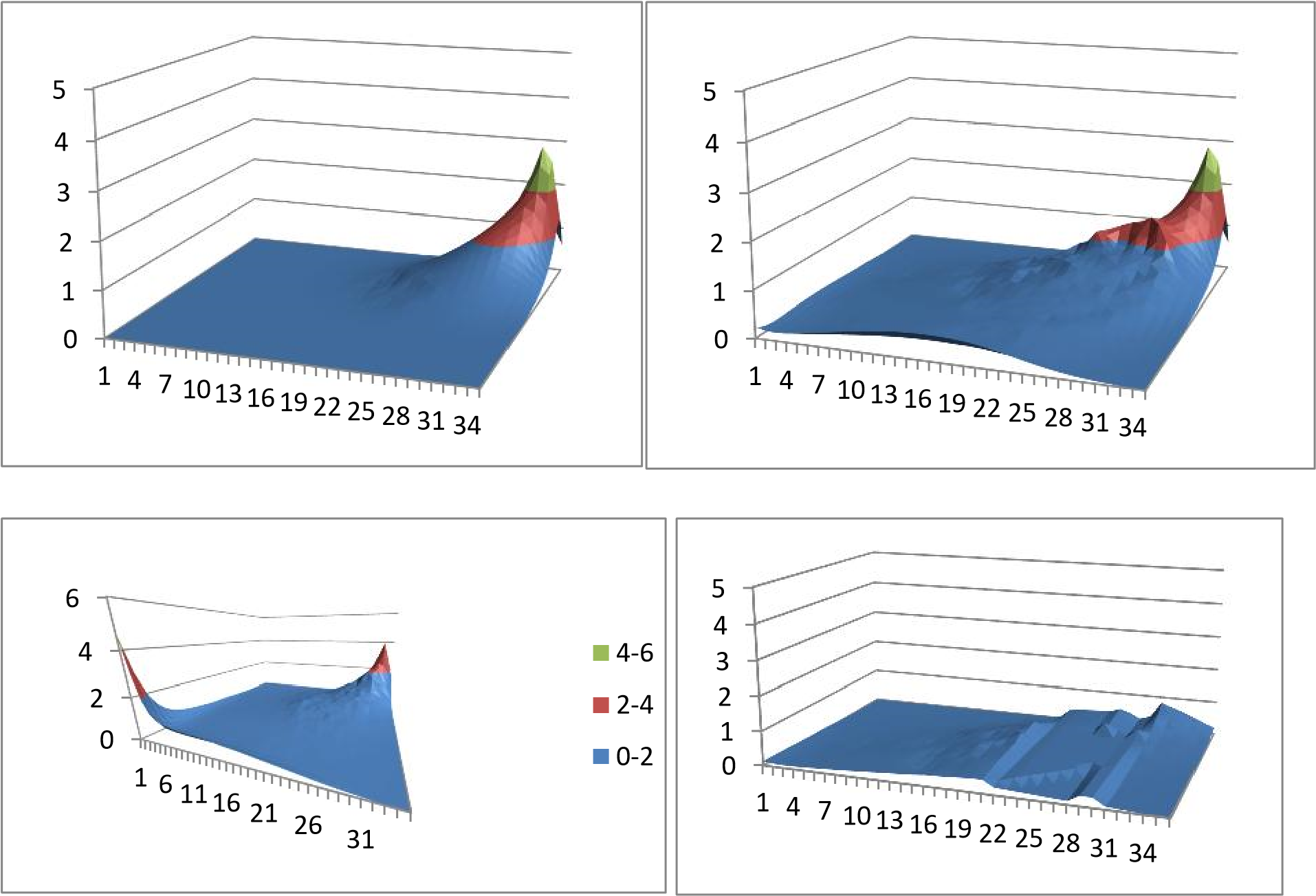}

\caption{ Movement of VEGF throughout the environment over the time period of angiogenesis (left to right, top to down)}
\end{figure}

\subsection{6.7 Completion of Cancer induced Angiogenesis (Criticality Detection)}
Based upon the factors that we test above, we come to the conclusion that the cellular density, the distance to the cell, and the overall density of the system are the most immediate and influential factors in angiogenesis. Given that successful angiogenesis results in heightened proximity between the tumor and the cardiovascular system, it has several implications for the future proliferation of the tumor. Relieved from hypoxia, the cancer will continue growing and possibly metastasize. Thus, we can successfully say that the chances of angiogenesis occurring at 102 um is 0\% (it is unnecessary) and also at 153 um (the upper limit).
Furthermore, the success of angiogenesis depends upon the cellular density and the number of cells in the region. We can factor this into the probability through a calculated factor combining the number of endothelial cells, and their surface area and volume. The sensitivity of this variable depends upon the overall density of the system, a weighted average of the level of extracellular proteins in the ECM. The effect of cellular density is compounded upon the distance of the tumor, and by integrating these factors, we can create an integrated probability model. With this probability model, we can rudimentarily predict the criticality of the tumor based upon the likelihood of angiogenesis occurring.
\section{Final Words}
The mathematical model developed in this work lays the theoretical foundation for practical cancer spread algorithms in the future. Turning away from typical models of cancerinduced angiogenesis, our project provides a refreshing view on cancer proliferation. Using the Cellular Potts model we not only consider more environment related factors, increasing accuracy, but these new factors can be assessed through patient tests. We establish a set of criteria for proliferation of a hypoxic tumor to occur; namely, the distance of the tumor from the nearest vasculature system, the cell density, and the concentration of interstitial fluid in the environment. Through these criteria, we develop a set of guidelines for determining criticality of a particular tumor.
\par
Our foray into modeling cancer angiogenesis through new lenses opens up many possible areas for new research. The development of our mathematical model now allows for practical applications in mapping cancer angiogenesis. Our model solves many of the previous problems that existed with previous continuous and discrete simulations. This model is computationally faster, which allows for quick repetitive simulations required in the analytical field. Furthermore, the use of the modified Cellular Potts system allows for varied input and manipulation of fundamental constants useful when dealing with diversified patients. Overall, the computational efficiency, input parameterization make this a model that serves well as a framework for practical investigations of cancer proliferation through cancer-induced angiogenesis.







\end{document}